\documentclass[aps]{revtex4}
\usepackage{graphicx}% Include figure files
\begin{document}
\title{Charged current neutrino induced coherent pion production}
\author{L. Alvarez-Ruso$^{a}$, L. S. Geng$^{a}$, S. Hirenzaki$^{b}$ and M. J. Vicente Vacas$^{a}$}
\affiliation{$^{a}$Departamento de F\'{\i}sica Te\'orica and IFIC,
Universidad de Valencia - CSIC;
Institutos de Investigaci\'on de Paterna, Aptdo. 22085, 46071 Valencia, Spain.\\
$^b$Department of Physics, Nara Women's University, Nara, 630-8506, Japan.}

%\date{\today}

\begin{abstract}
We analyze the neutrino induced charged current coherent pion production at the energies of interest for
recent experiments like  K2K and MiniBooNE. Medium effects in the production mechanism and the distortion of
the pion wave function, obtained solving the Klein Gordon equation with a microscopic optical potential, are included in the calculation. We find a strong reduction of the cross section due to these effects and also substantial modifications of the 
energy distributions of the final lepton and pion.
\end{abstract}

%\pacs{13.15+g,25.30Pt, 13.60Rj; 13.75Ev}
\maketitle

\section{Introduction}

Recent experimental results from the K2K collaboration show a  significant deficit of muons in the forward
scattering events with respect to the simulations, which could limit the accuracy of the predicted  neutrino energy spectrum at the far detector~\cite{Ahn:2002up}. This deficit could be due, among other possibilities, to the overestimation of
charged current (CC) coherent pion production.  A later search~\cite{Hasegawa:2005td}  found no evidence of CC
coherent pion production in $^{12}$C and obtained an upper limit for  the fraction of this process to the
total CC interaction well below some estimations based on the Rein and Sehgal model of Ref.~\cite{Rein:1982pf}.
Preliminary MiniBooNE results for neutral current $\pi^0$ production show  a similar deficit \cite{Raaf:2004ty} in the forward direction when compared with different MonteCarlo (MC)  models: NUANCE~\cite{Casper:2002sd},
NEUGEN~\cite{Gallagher:2002sf} and NEUT~\cite{Hayato:2002sd}. The discrepancies between these MC simulations indicate the considerable uncertainties in the theoretical description of coherent pion production. 
In addition, MiniBooNE has collected a large set of data for   
$\pi^+$ production in $^{12}$C induced by muon neutrinos of energies around 
0.7~GeV~\cite{Wascko:2006tx}. A fraction of these pions is created coherently, so that a  
realistic description of the coherent process is required to analyze and understand these 
data.

CC coherent pion production has been observed experimentally at higher energies and for several nuclei
\cite{Grabosch:1985mt,Marage:1986cy,Allport:1988cq,Aderholz:1988cs,Willocq:1992fv,Vilain:1993sf}. These
experiments were studied theoretically using models based on PCAC  \cite{Rein:1982pf,Belkov:1986hn} which
described well the results.  Work along the same lines has been carried out at low energies in Ref.
\cite{Paschos:2005km}. The only nuclear medium effect considered in these calculations is the distortion of
the final pion.

Other approaches have tried to incorporate some additional nuclear medium effects that modify the weak pion
production. In Ref.~\cite{Kim:1996az}, the authors used the impulse approximation with
undistorted pion waves, but already modified the  $\Delta$ resonance properties, and therefore the
production mechanisms, using an effective $\Delta$ mass. The importance of these nuclear effects was demonstrated there
by comparing the results with those obtained using a free $\Delta$. Kelkar et al.~\cite{Kelkar:1996iv} developed
a more sophisticated treatment of the $\Delta$ in the nuclear medium, and also included the final pion
distortion by solving the Klein Gordon (KG) equation with a pion nucleus optical potential. This model
predicted a very low cross section, compatible with the recent results of Ref.~\cite{Hasegawa:2005td}.
Nonetheless, there were several approximations in this work. On the one side a non relativistic reduction
of the hadronic current was done, on the other side all the transverse  parts of the amplitude were
neglected. Whereas these approximations are quite reasonable 
to get an estimate of the cross section
for this process, as discussed in Ref.
\cite{Kelkar:1996iv}, a more complete calculation is required now that new data are becoming available.  More
recently, Singh et al.~\cite{Singh:2006bm,Ahmad:2006cy} used similar medium effects on the production
mechanisms and improved on the description of the elementary $\nu+N\rightarrow N+\mu^-+\pi^+$ process by
using a fully relativistic calculation of this process and including all pieces of the amplitude.
However, the pion distortion was implemented in the eikonal approximation, which is not very reliable for
the low energy pions that apparently dominate this reaction. Our aim in this paper is to improve the model of Kelkar et
al.~\cite{Kelkar:1996iv} in a similar manner, using a more complete and relativistic elementary amplitude but still
keeping a more realistic treatment of the pion distortion which is calculated solving the KG equation.

In the following, we describe the formalism, including $\Delta$ production and decay, medium effects on the
production mechanism and the optical potential responsible for the distortion of the pion. In Section \ref{sec:re}, we present our results, compare with the available experimental data and make predictions for other
nuclei and observables. Finally, our summary and conclusions appear
 in Section \ref{sec:su}.

\section{Theoretical model}
\label{sec:th}

In the CC coherent pion production induced  by neutrinos ($\nu+A\rightarrow A+\mu^-+\pi^+$), the nucleus
remains in its ground state. The process consists of a weak pion production followed by the strong
distortion of the pion in its way out of the nucleus.  
 %Thus, it is suited to be studied using the Distorted Wave Born Approximation (DWBA). 
 First, we  must consider the elementary process of $\pi^+$ production
($\nu+N\rightarrow \mu^-+N+\pi^+$). There are large experimental uncertainties in this cross section at the low
energies relevant to this work, which can be clearly appreciated by comparing, 
for instance, Refs.~\cite{Radecky:1981fn} and \cite{Kitagaki:1986ct}.  
These discrepancies are ultimately translated into  different values for the $\Delta$ 
resonance axial form factors that appear in the theoretical models, and are a source of uncertainty
for the calculation of the pion production cross sections. In any case, it seems to be
clear that for neutrino  energies below 2 GeV, this process is dominated by  a
$\Delta(1232)$ excitation, $\nu+p\rightarrow \mu^-+\Delta^{++}$ or $\nu+n\rightarrow \mu^-+\Delta^{+}$ followed by
its decay~\cite{Fogli:1979cz,Kim:1996bt,Singh:1998ha,Alvarez-Ruso:1998hi,Sato:2003rq,Lalakulich:2005cs,Leitner:2006ww}, and
therefore it is larger for  protons than for neutrons as discussed below. The situation is different for
$\pi^0$ production or at higher energies. In that case, the $\Delta$ mechanism  is not enough to provide 
a good description of data~\cite{Leitner:2006ww}. 
After the introduction of the elementary model for the reaction we discuss how  it is modified in
the nuclear medium, due to the density dependent changes of the $\Delta$ resonance peak's position and width. Finally, we 
study the distortion of the pion wave function using a pion nucleus optical potential.

\subsection{$\Delta$ production and decay}
The matrix element for the elementary process 
$\nu_\mu(k)+n(p) \rightarrow \Delta ^{+}(p')+ \mu^-(k')$
is written as 

\begin{equation}
\label{eq:1}
{\cal M}_{n,\Delta^+}= {{G}\over{\sqrt{2}}}\, \cos \theta_{c}\, l_{\alpha} J^{\alpha}_{\Delta}\,,
\end{equation}
with the leptonic current

\begin{equation}
 l_{\alpha}= \bar{u}_\ell(k') \gamma_{\alpha} (1 -\gamma_{5}) u_{\nu_\ell}(k)\,,
\end{equation}
and the hadronic current
\begin{eqnarray} 
 J^{\alpha}_{\Delta}=\bar{\psi}_{\mu}(p') {\cal A}^{\mu\alpha} u(p)\,,
\end{eqnarray}  
\begin{eqnarray} 
\nonumber {\cal A}^{\mu\alpha}=&\{ 
  {{C_3^V}\over{M}} (g^{\mu \alpha} q \!\!\! / - q^{\mu} \gamma^{\alpha})+
  {{C_4^V}\over{M^2}} (g^{\mu \alpha} q\cdot p' - q^{\mu} p'^{\alpha})  
  + {{C_5^V}\over{M^2}} (g^{\mu \alpha} q\cdot p - q^{\mu} p^{\alpha})
   \}\gamma_{5}\\ 
  &+\{ {{C_3^A}\over{M}} (g^{\mu \alpha}q \!\!\! / - q^{\mu} \gamma^{\alpha})+
 {{C_4^A}\over{M^2}} (g^{\mu \alpha} q\cdot p' - q^{\mu} p'^{\alpha})+
 {C_5^A} g^{\mu \alpha}+ {{C_6^A}\over{M^2}} q^{\mu} q^{\alpha} \}\,,
\end{eqnarray}  
where $M$ is the nucleon mass, $\psi_{\mu}(p')$ and $u(p)$ are the 
Rarita Schwinger and Dirac spinors for the $\Delta$ and the nucleon of momentum $p'$ 
and $p$, $q=p'-p=k-k'$ is the momentum transfer, $C_i^V$ and $C_i^A$ 
($i=3,4,5,6$) are the vector and axial vector transition form factors.
The amplitude for the process on protons,  
$\nu_\mu(k)+p(p) \rightarrow \Delta ^{++}(p')+ \mu^-(k')$, is related to the previous one
by an isospin factor
\begin{equation}
{\cal M}_{p,\Delta^{++}}=\sqrt{3}{\cal M}_{n,\Delta^+}\,.
\end{equation}

The vector form factors can be related to the electromagnetic ones.
The conservation of the vector part of the current implies that $C_6^V=0$. The assumption of $M_{1+}$ dominance 
for the $\Delta$ electroproduction amplitude gives~\cite{Fogli:1979cz}
\begin{eqnarray}
C_5^V=0,\; C_4^V=-\frac{M}{M_\Delta} C_3^V\,,
\end{eqnarray}
and for $C_3^V$ we take \cite{Schreiner:1973mj}
\begin{eqnarray}
C_3^V={{2.05}\over{(1-q^2/0.54\,{\rm GeV}^2)^2}}\,.
\end{eqnarray}

Except for $C_6^A$ that can be related to $C_5^A$ using PCAC, there are no other constraints  for the axial 
form factors. We use the following ones, fitted to neutrino scattering data
\cite{Bijtebier:1970ku,Zucker:1971hp,Schreiner:1973mj,Barish:1978pj,Bell:1978rb,Allen:1980ti,
Radecky:1981fn,Kitagaki:1990vs,Alvarez-Ruso:1997jr,Alvarez-Ruso:1998hi}
\begin{equation}
C_{i=3,4,5}^A(q^2) = {{C_i^A(0)\left[ 1-{{a_i q^2}\over{b_i-q^2}} \right] }
{\left( 1- {{q^2}\over{M_A^2}}\right)^{-2}}} ,
\end{equation}
and
\begin{equation}
C_6^A(q^2) = C_5^A {{M^2}\over{m_{\pi}^2-q^2}}\,,
\end{equation}
with $C_3^A(0)=0$, $C_4^A(0)=-0.3$, $C_5^A(0)=1.2$, $a_4=a_5=-1.21$,
$b_4=b_5=2$ GeV$^2$ and for the axial mass we take $M_A=1.28$~GeV~\cite{Kitagaki:1990vs}.
This set of form factors produces a good agreement with weak $\Delta$ production data
induced by  neutrinos on nucleons~\cite{Alvarez-Ruso:1998hi}.

A new analysis of world electron scattering data~\cite{Tiator:2003uu} allows to go beyond the 
$M_{1+}$ approximation and update the vector form factors~\cite{Lalakulich:2006sw}. 
This information calls for a new extraction of the axial form factors. 
Some steps in this direction have been recently taken in Refs.~\cite{Sato:2003rq,Hernandez:2007qq},
using  models for the elementary reaction which include, apart from $\Delta$ 
excitation, some background terms. 
For the sake of consistency with our description of pion production on the nucleon explained previously,
based on the $\Delta$ dominance, we stick to the set of form factors given above.

We still need to take into account the $\Delta$ decay into a pion and a nucleon. We use the following 
Lagrangian to describe the $\Delta N\pi$ transition~\cite{Benmerrouche:1989uc}
\begin{equation}
{\cal L}_{\Delta\pi N}=   \frac{f^*}{m_\pi}\bar{\Psi}_{\mu}\vec{T}^\dagger\cdot\partial^\mu\vec{\Phi}\,\Psi+h.c.\,,
\end{equation}
 where $\vec{T}^\dagger$ is the isospin 1/2 to 3/2 transition operator and 
${\Psi}_{\mu}$, $\vec{\Phi}$ and $\Psi$ are the $\Delta$, pion and nucleon fields respectively. The isospin operator produces a factor 1 for the decay  
$\Delta^{++}\rightarrow \pi^++p$
and a factor $1/\sqrt{3}$ for the decay $\Delta^{+}\rightarrow \pi^++n$. This,
together with Eq. (5), implies that the amplitude for CC $\pi^+$ production on
the proton is three times larger than on the neutron.
The coupling constant $f^*=2.13$ is such that the experimental $\Delta\rightarrow \pi N$ width is reproduced.
At the high $\Delta$ invariant masses reached with the K2K or MiniBooNE neutrino energies, 
the finite size of the hadrons becomes relevant. 
This can be empirically taken into account with a form factor $F(p')$ that modifies the 
$\Delta N\pi$ coupling. Here, we adopt the following ansatz
\begin{equation}
F(p')=\frac{\Lambda^4}{\Lambda^4+(p'^2-M_\Delta^2)^2},
\end{equation}
with $\Lambda=1$ GeV, used in coupled channel studies of pion and photoproduction of baryonic 
resonances~\cite{Penner:2002ma}.
The new hadronic current which should replace $J^\alpha_\Delta$ in Eq.~(\ref{eq:1}), already
incorporating the decay into a pion and a nucleon,  is given 
(for the case of $\Delta^+$) by
\begin{equation}
J^{\mu}_{N\pi}= -\frac{1}{\sqrt{3}} \frac{f^*}{m_\pi} p_\pi^\alpha F(p')
\bar{u}(p_f) D(p')\Lambda_{\alpha \beta}{\cal A}^{\beta\mu} u(p)\,,
\end{equation}
where $p_\pi$ and $p_f$ are the pion and final nucleon momenta so that $p'=p_\pi+p_f$. The $\Delta$ propagator is given by
\begin{equation}
\label{eq:delprop}
 D(p')=\frac{1}{(W+M_\Delta)(W-M_\Delta+i \Gamma_\Delta/2 )}\,,
\end{equation}
where $W=\sqrt{p'^2}$. The energy dependent $\Delta$ width is 
\begin{equation}
\Gamma_\Delta=\frac{1}{6\pi}\left(\frac{f^*}{m_\pi}\right)^2 F(p')^2 \frac{M_\Delta}{W}p_{\pi,cm}^3\,,
\end{equation}
with $ p_{\pi,cm}$ the pion momentum in the $\Delta$ rest frame.
Finally, the spin $3/2$ projection operator is given by
\begin{equation}
\Lambda_{\alpha \beta}=- \left(p'\!\!\!\! / + M_\Delta \right) \left( g_{\alpha \beta} - \frac{2}{3} \frac{p'_{\alpha } p'_{\beta }}{M_\Delta^2} 
     + \frac{1}{3} \frac{p'_{\alpha } \gamma_{\beta} - p'_{\beta } \gamma_{\alpha}}{M_\Delta } -
     \frac{1}{3} \gamma_{\alpha} \gamma_{\beta} \right).
\end{equation}

\subsection{$\Delta$ in the nuclear medium}
\label{sect:22}
The $\Delta$ properties are strongly modified inside the nuclear medium and have been the subject of intensive
study, both experimental and theoretical, for many years, see i.e. 
\cite{Kisslinger:1973np, Hirata:1978wp, Oset:1979tk, Horikawa:1980cv, Oset:1981ih, Freedman:1982yp, Hirata:1982mz, Oset:1987re} and
references therein. In this work we  use the results from Refs.~\cite{Oset:1987re,Nieves:1991ye}
where the $\Delta$ selfenergy is calculated in a many body approach as a function of the local baryon density
$\rho(r)$.
This model has been
extensively tested in pion induced inclusive processes~\cite{Salcedo:1987md,VicenteVacas:1993bk},
photonuclear reactions~\cite{Carrasco:1989vq,Carrasco:1991mb}, electron scattering~\cite{Gil:1997bm}
and even for coherent pion production induced by photons, electrons
or nuclei~\cite{Carrasco:1991we,Hirenzaki:1993jc,FernandezdeCordoba:1992ky,FernandezdeCordoba:1994wn}.
In the nuclear medium 
the $\Delta$ resonance acquires a selfenergy because of several effects such as Pauli blocking of the
final nucleon  and  absorption processes: $\Delta N\rightarrow NN$, 
$\Delta N\rightarrow NN\pi$ or $\Delta NN\rightarrow NNN$. The real part can be parametrized as
\begin{equation}
\mathrm{Re}\Sigma_\Delta(\rho)=\mathrm{Re}\Sigma^0_\Delta(\rho)+\frac{4}{9}\left(\frac{f^*}{m_\pi}\right)^2 g' \rho 
\approx 40\, \mathrm{MeV}\, \frac{\rho}{\rho_0}\,,
\end{equation}
where $\rho_0=0.17\,\mathrm{fm}^{-3}$ is the normal nuclear density. Here, in addition to the attractive proper selfenergy
$\mathrm{Re}\Sigma^0_\Delta(\rho)$,
the effective repulsive  contribution,  that comes from the iterated $\Delta$-hole excitation driven by the
Landau Migdal interaction with $g'=0.63$, has been added~\cite{Carrasco:1989vq}.
The imaginary part is parametrized by the expression
\begin{equation}
-\mathrm{Im}\Sigma_\Delta(\rho)=C_Q\left(\frac{\rho}{\rho_0}\right)^{\alpha}+C_{A2}\left(\frac{\rho}{\rho_0}\right)^{\beta}+
C_{A3}\left(\frac{\rho}{\rho_0}\right)^{\gamma}\,,
\end{equation}
where the terms with the coefficients $C_Q$, $C_{A2}$ and $C_{A3}$ correspond to the processes 
$\Delta N\rightarrow NN\pi$, $\Delta N\rightarrow NN$ and $\Delta NN\rightarrow NNN$ respectively.
The values of $C_Q$, $C_{A2}$, $C_{A3}$, $\alpha$, $\beta$ and $\gamma$ can be found in Eq. (4.5) and 
Table 2 of Ref.~\cite{Salcedo:1987md}. The parameterizations are given as a function of the
kinetic energy in the laboratory system of a pion that would excite a $\Delta$ with the corresponding invariant mass,
and are valid in the range 85 MeV $< T_\pi<$ 315 MeV. Below 85 MeV the contributions 
from $C_Q$ and $C_{A3}$ are rather small and are taken from \cite{Nieves:1991ye}, where the model was extended
to low energies. The term with $C_{A2}$ shows a very mild energy dependence and we still use the
parameterization from Ref.  \cite{Salcedo:1987md} even at low energies.
For $T_\pi$  above 315 MeV we have kept these selfenergy terms constant and equal to their values at the bound.
The uncertainties in these pieces are not very relevant there because the $\Delta \rightarrow N\pi$ decay becomes 
very large and absolutely dominant.
Finally, the Pauli blocking of the $\pi N$ decay reduces the $\Gamma_\Delta$ free width which now reads as
\begin{equation}
\Gamma_\Delta^{\mathrm{Pauli}}=\Gamma_\Delta \frac{I_1+I_2}{2}\,.
\end{equation}
The angular integrals $I_1$ and $I_2$ can be found in  Appendix B of Ref. \cite{Nieves:1991ye}.
This selfenergy is taken into account by making   the substitutions $M_\Delta\rightarrow M_\Delta
+\mathrm{Re}\Sigma_\Delta$ and $\Gamma_\Delta/2\rightarrow\Gamma_\Delta^{\mathrm{Pauli}}/2-\mathrm{Im} \Sigma_\Delta$ in the propagator of Eq.~(\ref{eq:delprop}).

\subsection{Cross section}
The hadronic current is further modified in the nucleus, where the nucleons are bound and thus have a
momentum distribution. In the impulse approximation, and after summing over all nucleons, we can write it as 
\begin{equation}
J^{\mu}_{N\pi}= - \frac{i}{2}\int d^3r e^{i(\vec{q}-\vec{p}_\pi)\cdot \vec{r}}
\left[\rho_p(r)+\frac{\rho_n(r)}{3}\right]
\sqrt{3}\frac{f^*}{m_\pi} F(p') \tilde{D}(p') p_\pi^\alpha 
\mathrm{Tr}\left\{\bar{u}(0) \Lambda_{\alpha \beta}{\cal A}^{\beta\mu} u(0)\right\}\,,
\end{equation}
where $\tilde{D}(p')$ is the in-medium $\Delta$ propagator and the trace corresponds to the sum over the nucleons' spin.
In the evaluation  of the trace we have taken an average nucleon momentum for the spinors neglecting corrections of order
$(p/M)^2$.
Given the large mass of
the nucleus, the pion energy is taken to be equal to $q^0=E_\nu-E_\mu$ and this defines the
modulus of the asymptotic pion momentum. On the other hand, a momentum of $\vec{q}-\vec{p}_\pi$ is
transferred to the nucleus. In the evaluation of the amplitude we assume that this momentum is equally shared
by the initial and the final nucleon, so that they have $\vec{p}_i=(\vec{p}_\pi-\vec{q})/2$ and 
$\vec{p}_f=(\vec{q}-\vec{p}_\pi)/2$ respectively. 
This prescription has also been used in Refs.~\cite{Carrasco:1991we,Drechsel:1999vh}  for coherent $\pi^0$ photo- and electroproduction. The approximation  is based on the fact that, for Gaussian nuclear wave functions, it leads to an exact treatment of the terms linear in momentum of the elementary amplitude and
allows for a consistent description of the pion-nucleon and pion-nucleus kinematics~\cite{Drechsel:1999vh}.
Pion distortion is taken into account by the replacements
\begin{equation}
 e^{-i\vec{p}_\pi\cdot\vec{r}} \;\rightarrow\;  \phi^*_{out}(\vec{p}_\pi,\vec{r})\,,
\end{equation}
and
\begin{equation}
\vec{p}_\pi e^{-i\vec{p}_\pi\cdot\vec{r}} \;\rightarrow\;i \vec{\nabla} \phi^*_{out}(\vec{p}_\pi,\vec{r})\,,
\end{equation}
where $\phi^*_{out}(\vec{p}_\pi,\vec{r})$ is an outgoing  solution of the KG equation for a pion of
asymptotic momentum $\vec{p}_\pi$ calculated with the optical potential described below.
To test the validity of the eikonal approximation we have also calculated the cross section with the pion wave function
\begin{equation}
 \phi^{*,eik}_{out}(\vec{p}_\pi,\vec{r})=e^{-i\vec{p}_\pi\cdot\vec{r}}
 e^{-i\int^\infty_z \frac{\Pi(\rho(\vec{b},z'))}{2p_\pi}dz'}\,,
\end{equation}
where $\vec{r}=(\vec{b},z)$ and $\Pi$ is the pion selfenergy described in the next section.

The  cross section for the coherent process $\nu+A\rightarrow A+\mu^-+\pi^+$ is then given by
\begin{equation}
\frac{d \sigma}{d \Omega_\ell d E_\ell d \Omega_\pi} = \frac{1}{8}
\frac{|\vec{k}'| |\vec{p}_\pi|} {|\vec{k}|} \frac{1}{(2 \pi)^5} \,
|{\cal M} |^2\,,
\end{equation}
with
\begin{equation}
{\cal M}= {{G}\over{\sqrt{2}}}\, \cos \theta_{c}\, l_{\alpha} J^{\alpha}_{N\pi}\,.
\end{equation}

\subsection{Pion optical potential and distorted wave function}

The pion wave function is the solution of the KG equation with an optical potential. Since
most of the produced pions lie in the energy region around the $\Delta$ excitation  the $\Delta$-hole
model can be used.
%We have also explored the impact of other contributions important at low energies, like
%those related to nucleon-hole excitations or s-wave $\pi N$ scattering. However, they only affect a very small region
%of the phase space and  would produce minor corrections. 
In this model, which we take from Ref.~\cite{Oset:1987re}, the optical potential is given by
\begin{equation}
V_{opt}=\frac{\Pi}{2\omega}\,,
\end{equation}
where $\omega$ is the pion energy in the laboratory system and the selfenergy $\Pi$ reads as
\begin{equation}
\label{eq:eq}
\Pi=-4\pi\frac{M^2}{s}\vec{q}\,^2 \frac{\cal P}{1+4\pi g'{\cal P} } \,.
\end{equation}
Here, $s$ is the Mandelstam variable of the $\pi N$ system and $\vec{q}$ is the pion laboratory momentum. Finally,  ${\cal P}$ is given by
\begin{equation}
{\cal P}=-\frac{1}{6\pi}\left(\frac{f^*}{m_\pi}\right)^2 
\left \{
\frac{\rho_p+\rho_n/3}{\sqrt{s}-M_\Delta-\mathrm{Re}\Sigma^0_\Delta+i \Gamma_\Delta^{\mathrm{Pauli}}/2- i\,\mathrm{Im} \Sigma_\Delta}+
\frac{\rho_n+\rho_p/3}{-\sqrt{s}-M_\Delta+2M-\mathrm{Re}\Sigma^0_\Delta}
\right \}\,,
\end{equation}
where the two terms correspond to direct and crossed $\Delta$-hole excitations.
The neutron densities $\rho_n$ are taken from Ref.~\cite{Garcia-Recio:1991wk} and the
proton densities $\rho_p$ are obtained from the parameterizations compiled in
Ref.~\cite{DeJager:1987qc}, both deconvoluted
 to take into account the finite size of the nucleons as done in 
\cite{Garcia-Recio:1991wk}. The real and imaginary part of the $\Delta$ selfenergy $\Sigma_{\Delta}$ and the
Pauli corrected width, $\Gamma_\Delta^{\mathrm{Pauli}} $, have been described in section~\ref{sect:22}. 

In coordinate space and for finite nuclei, this p-wave potential can be cast as
\begin{equation}
2 \omega \,V_{opt} (\vec{r}) = 4 \pi \frac{M^2}{s}\left[ \vec{\nabla} \cdot \frac{{\cal P}(r)}{1+4\pi g' {\cal P}(r)} \vec{\nabla} 
- \frac{1}{2}\frac{\omega}{M} \Delta \frac{{\cal P} (r)}{1+4\pi g'{{\cal P} (r)}}  \right] \,,
\end{equation} 
where the first term has the standard Kisslinger form and the second one accounts for the angular transformation from 
center-of-mass to laboratory variables. The $r$ dependence in $\cal P$ appears via the local density approximation 
$\rho \rightarrow \rho (r)$. 
With this potential we solve the KG equation and obtain the pion scattering wave function.  The procedure is described in detail in
Refs.~\cite{Nieves:1991ye, Nieves:1993ev}. To asses the quality of the potential  one can compare 
its results for differential cross sections with  pion nucleus
elastic  scattering data. We obtain an overall good agreement from light to heavy nuclei at the energies relevant for this work. We
can also refer the reader to  the article by Garcia Recio et al.~\cite{Garcia-Recio:1989xa}, where a very similar potential was
considered, using as here the local density approximation  and the same values for the $\Delta$ selfenergy. The main differences
with that work are their inclusion of a  phenomenological s-wave selfenergy, and our inclusion of the $\Delta$ crossed term. We
have checked that both pieces produce  only minor effects in the  $\Delta$ resonance region.

\section{Results}
\label{sec:re}
\begin{figure}
       \centering
	\includegraphics[width=0.82\textwidth]{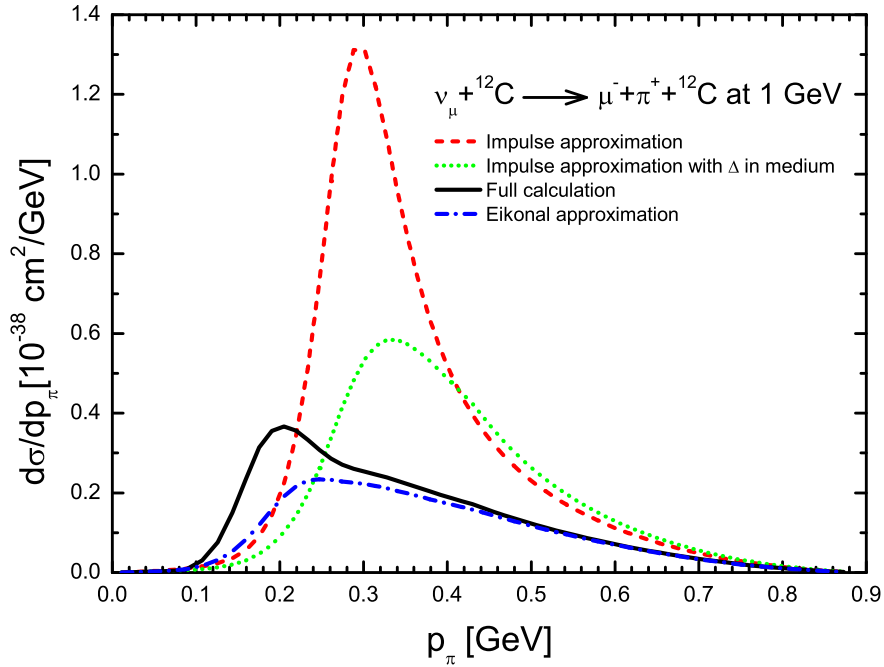}
	\caption{(Color online) Momentum distribution of the coherent pions.}
	\label{fig:fig1}
\end{figure}
We show in Fig. \ref{fig:fig1} the differential cross section $d\sigma /dp_{\pi}$ for CC coherent pion production in $^{12}$C at a neutrino energy $E_{\nu}=1$ GeV. This plot shows the effect of the different ingredients of the calculation.
The modification of the elementary reaction mechanism, through the inclusion of the $\Delta$ selfenergy in the propagator,
already produces a strong reduction of the cross section. This reduction of around 35\% agrees with the results of Ref.
\cite{Singh:2006bm}. 
 The pion distortion further  decreases 
the cross section and moves the peak to lower energies. This reflects the presence of a strongly absorptive part in the optical potential around the $\Delta$ resonance peak.  The final result shows that the pion spectrum is peaked at much lower energies or, equivalently, that the muon energy distribution is peaked at higher energies
than for the impulse approximation. As expected, the eikonal approximation fails for low and intermediate energies, where a better treatment of the pion wave function is clearly required. Our eikonal result differs from the one obtained in Ref.~\cite{Singh:2006bm} because of their use of the asymptotic momentum in the amplitude, whereas we take the gradient 
of the distorted pion wave function, and our more complete treatment of the optical potential (see Eq.~(\ref{eq:eq}) vs. Eq.~(6) of  Ref.~\cite{Singh:2006bm}).

\begin{figure}
       \centering
	\includegraphics[width=0.82\textwidth]{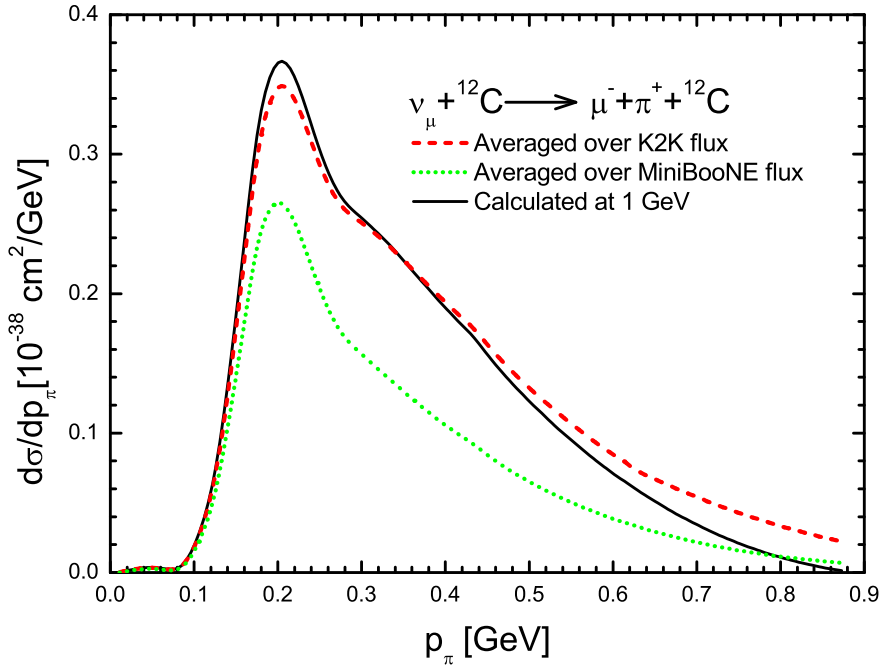}
\caption{(Color online) Momentum distribution of the coherent pions averaged over the K2K and MiniBooNE fluxes; the same distribution at  $E_\nu=1$~GeV is also plotted for reference.}
	\label{fig:fig2}
\end{figure}
In Fig. \ref{fig:fig2}, we present the same observable averaged over the K2K~\cite{Ahn:2006zz} and the MiniBooNE \cite{Monroe:2004xe} spectra compared with the results for a fixed neutrino energy. The consideration of high energy neutrinos widens the pion momentum distribution, due to phase space. Even when both spectra are quite different and 
K2K has a larger average neutrino energy (1.3~GeV) than MiniBooNE (0.75~GeV), the peak position stays at the same pion momentum, below the $\Delta$ resonance. Furthermore,
most of the pions have relatively low energies such that the use of the $\Delta$-hole model is appropriate. 
 
\begin{figure}
       \centering
	\includegraphics[width=0.99\textwidth]{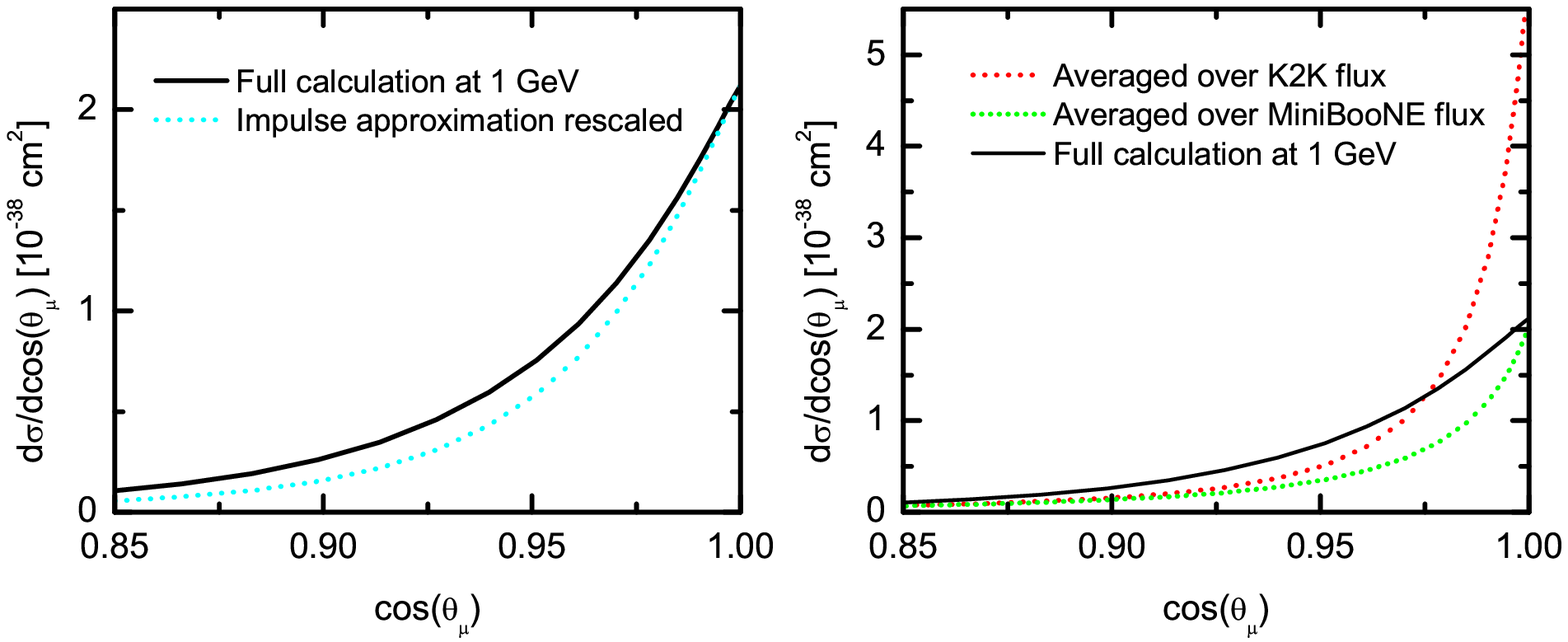}
\caption{(Color online) Muon angular distribution  in CC coherent pion production. Left panel: full model vs. impulse approximation. The latter has been rescaled to match the full calculation at zero degrees. Right panel: full model  at $E_\nu=1$ GeV   and averages over K2K and MiniBooNE fluxes.}
	\label{fig:fig3}
\end{figure}

 Although the total cross section and the energy distribution are strongly modified by the nuclear medium, the angular distribution of the muons remains relatively unaffected, as can be seen on the left panel of Fig.~\ref{fig:fig3} where we compare the result of our model with the impulse approximation, rescaled to match the full calculation at zero degrees. On the right panel, the muon angular distributions averaged over the K2K and the MiniBooNE neutrino spectra are shown together with the one obtained for 1~GeV neutrinos. 
In both cases, the consideration of higher energy neutrinos leads to a narrower angular distribution.
As a consequence, the MiniBooNE angular distribution is appreciably more forward peaked than the one at 1~GeV even when its average energy is lower.    
 We have also checked that the lower bound on the muon energy used by K2K in Ref.~\cite{Hasegawa:2005td} ($p_{\mu} > 450$~MeV/c) does not modify appreciably this observable.

\begin{figure}
       \centering
	\includegraphics[width=0.82\textwidth]{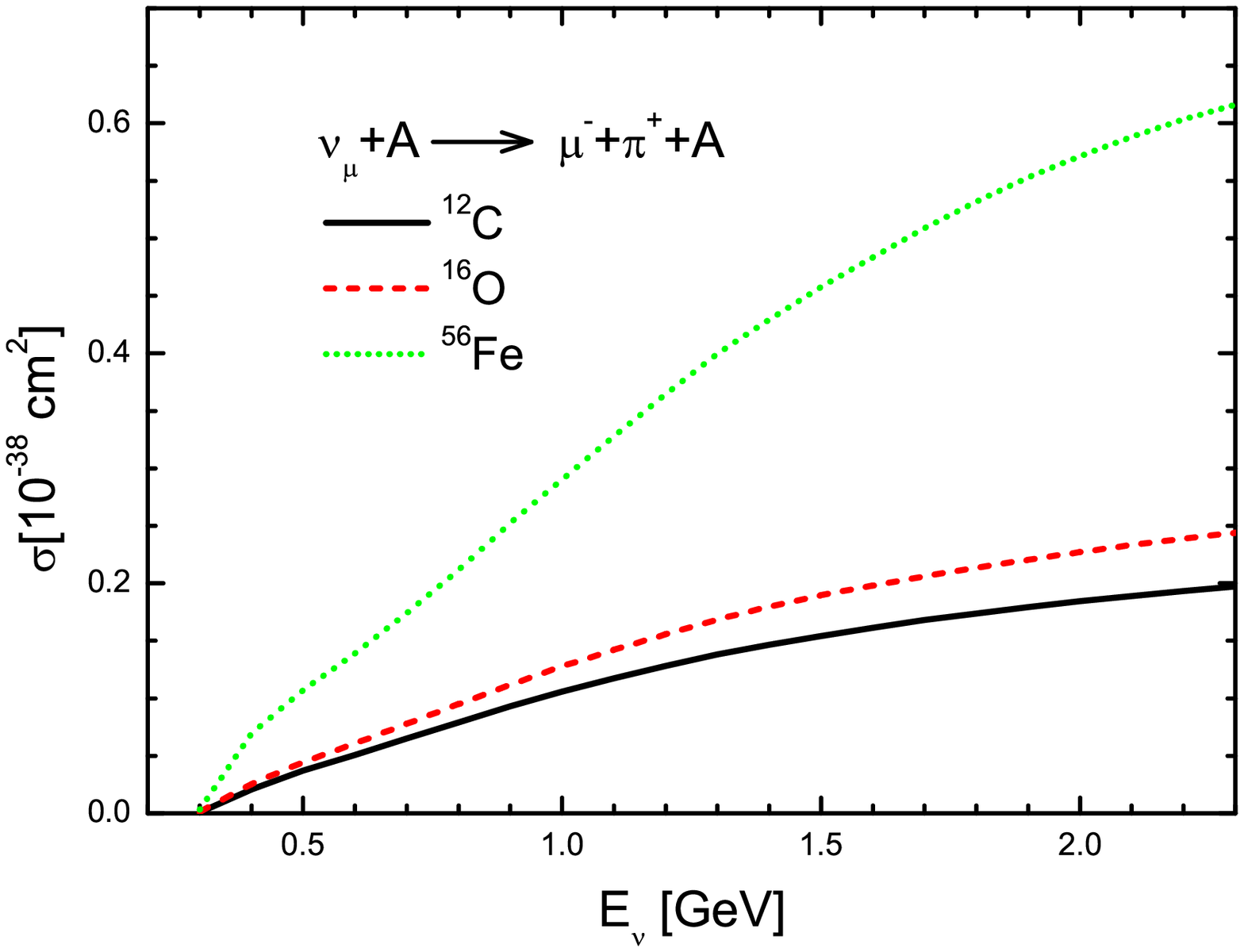}
\caption{(Color online) Total cross section for CC coherent pion production as a function of the neutrino energy for several nuclei.}
	\label{fig:fig4}
\end{figure}
We show the total cross section as a function of the neutrino energy and for several nuclei in  Fig.~\ref{fig:fig4}, with the caveat that the pion production model is less satisfactory at high energies. One reason is that mechanisms, other than 
the excitation of the $\Delta$ resonance, become relevant. 
Also the pion distortion, based on the 
$\Delta$-hole model, is not appropriate for the  high energy pions that can be produced by  neutrinos with $E_\nu > 2 $~GeV.
Nonetheless, the $\pi N$ interaction is much weaker at high energies than at the $\Delta$ peak and thus distortion effects should be smaller there.

\begin{figure}
       \centering
	\includegraphics[width=0.82\textwidth]{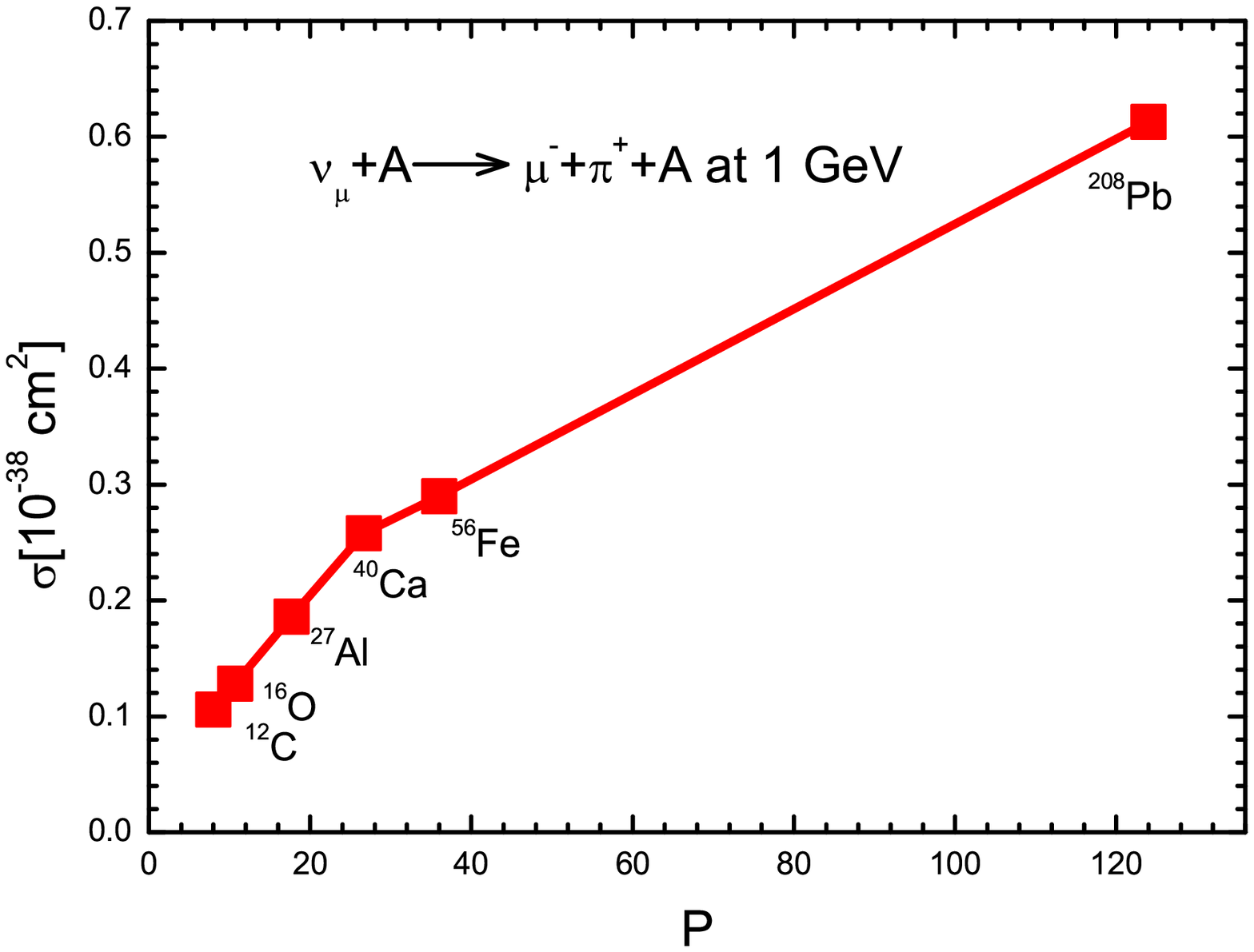}
	\caption{(Color online) Total cross section  for CC coherent pion production at $E_\nu=1$~GeV as a function of the effective number of participants P=Z+N/3, as explained in the text.}
	\label{fig:fig5}
\end{figure}

In Fig. \ref{fig:fig5}, we present the dependence of the total cross section on the atomic number. 
In this process, the amplitude is the coherent sum of the contributions of all participant nucleons. Taking into account the isospin factors, this implies that the amplitude is proportional to an effective number of participants defined as P=Z+N/3; here Z and N are the number of protons and neutrons respectively. 
This could suggest a quadratic dependence of the cross section on P, so that for heavier nuclei,
the process could be comparatively larger with respect to incoherent $\pi$ production or other processes.
However, there are several reasons that quench the P dependence. First, pion absorption is quite strong and forces the reaction to be peripheral.
Notice that the  cross sections for the inclusive processes are not affected by this. 
Second, the nuclear form factor is narrower for heavy nuclei, and reduces more the contribution from high momentum transfers.

Our result for the integrated cross section for coherent pion production on 
$^{12}$C averaged over the K2K flux is $\sigma^{CC}_{coh}=10. \times 10^{-40}$~cm$^2$.
This value is above the upper limit of $7.7 \times 10^{-40}$~cm$^2$ obtained
using the ratio between coherent and $\sigma^{CC}$, the total CC cross section, from the K2K collaboration~\cite{Hasegawa:2005td} and the value for $\sigma^{CC}$ of their  MC calculation. 
Without the experimental threshold for the muon momentum, $p_\mu>450$ MeV$/c$, we obtain $\sigma^{CC}_{coh}=12. \times 10^{-40}$~cm$^2$.

\begin{figure}
       \centering
	\includegraphics[width=0.82\textwidth]{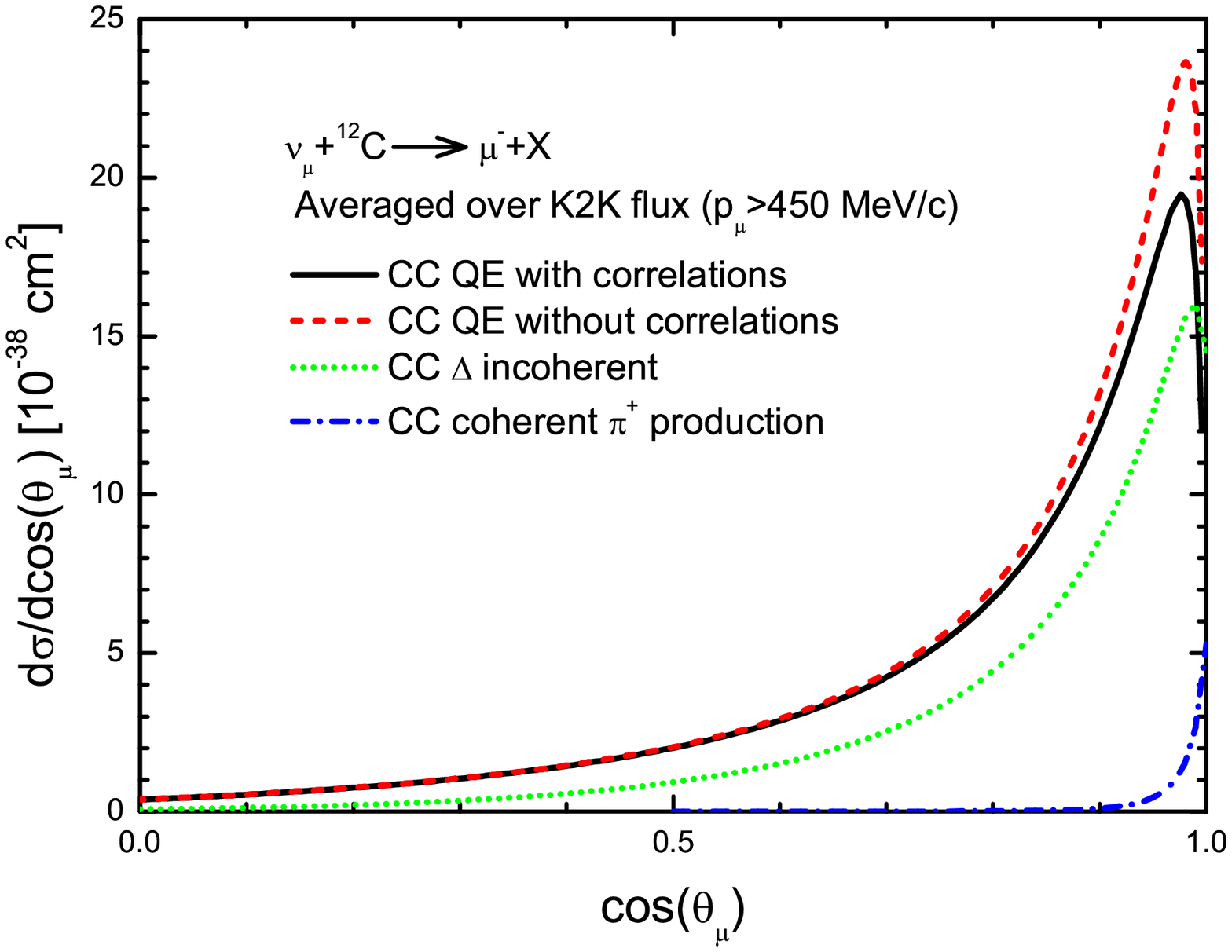}
	\caption{(Color online) Muon angular distribution averaged over the K2K flux for incoherent CC quasielastic (QE) scattering, 
$\Delta$ excitation and coherent $\pi^+$ production.}
	\label{fig:fig6}
\end{figure}
There are several factors that might conspire to produce the disagreement between our 
calculation and the experimental upper bound. First of all, due to the 
considerable uncertainty in 
the experimental data for pion production cross sections on the nucleon, the axial N-$\Delta$ form 
factors are not sufficiently constrained. A more complete theoretical description of the elementary 
amplitude with the inclusion of background terms~\cite{Fogli:1979cz,Hernandez:2007qq}
and heavier resonances~\cite{Fogli:1979cz,Lalakulich:2006sw} could help, 
but in order to put such a model on a firm ground, more precise data are required. The optical potential employed in our calculation is realistic around the $\Delta$ peak (i.e. for pions with 150-450~MeV/c momenta), where most 
of the strength of the reaction actually concentrates, but has room for improvement both at 
lower and higher energies. 

It is also important to recall that at forward angles, where the 
coherent process is sizable, nuclear effects play an important role and this may affect the 
experimental separation of the coherent events from the incoherent ones which,
to a large extent, relies on the theoretical models built in the MC simulations.
The situation is illustrated in Fig.~\ref{fig:fig6} where we plot the muon angular 
distributions averaged over the K2K flux (and with $p_\mu > 450$~MeV/c)
for coherent $\pi^+$ production, together with the main contributions to the total 
inclusive CC cross section: quasielastic scattering (QE) and incoherent $\Delta$ excitation. 
The calculations of the $\Delta$ part is performed with the same elementary amplitudes and 
in-medium effects as given above for the coherent reaction. For the quasielastic process, we have adopted the model 
of Ref.~\cite{Singh:1992dc} but updating the nucleon form factors according to~\cite{Budd:2003wb}. 
Nuclear effects include Fermi motion, Pauli blocking with a local Fermi gas and
the renormalization of the weak transition, which is treated as an  
RPA resummation of particle-hole and $\Delta$-hole states. These nuclear correlations cause a considerable reduction 
of strength at low $q^2$ (forward angles), as can be seen in Fig.~\ref{fig:fig6}, while they are negligible for
$\cos\theta_\mu < 0.8$. Therefore, if a model that lacks these correlations is used to extrapolate the data from the 
region of $\cos\theta_\mu \lesssim 0.8$ to forward angles, one might overestimate the QE part, causing an 
underestimation of the contribution of other mechanisms, like the coherent pion production, to the cross section.

\section{Summary and Conclusions}
\label{sec:su}
We have studied CC coherent pion production induced by muon neutrinos $\nu_\mu + A \rightarrow \mu^- + \pi^+ + A$. Our model
takes into account the modification of the production mechanism due to the renormalization of the $\Delta$ properties in the nuclear medium and the distortion of the final pion. The distorted pion wave function is obtained by solving the Klein Gordon equation with an optical potential based on the $\Delta$-hole model. Both effects produce a large reduction of the cross section with respect to the impulse approximation. The distortion of the pion shifts the peak of the pion energy distributions towards low energies.  The angular distributions are slightly widened by the nuclear medium effects. 
 
The A dependence of the coherent process has been investigated. We have found that the integrated cross section grows more slowly than one would naively expect due to the strong pion absorption and the effect of nuclear form factors. Therefore, we do not expect that coherent pion production becomes more relevant with respect to the incoherent processes for heavier nuclei.   

We have also studied the cross sections averaged over the K2K and MiniBooNE spectra. In the case of K2K, we find a cross section 30\% larger than the upper limit estimated in the experiment. One should however remember that such upper limit is based on values for the CC cross sections on nuclei (quasielastic, pion production, etc.)  which are not well known at the rather low energies discussed here. 
Further improvements in the theoretical description of this reaction will require more precise data for the neutrino nucleon
cross sections at low energies that could constrain the axial N-$\Delta$ form factors.
As for MiniBooNE, we give predictions for energy and angular distributions
which can be  useful to compare with and analyze their data.

\begin{acknowledgments}
We thank P. Novella and J. J. Gomez-Cadenas for useful discussions. This work was partially supported by DGI and FEDER funds,
contract  BFM2003-00856, by the EU Integrated Infrastructure
Initiative Hadron Physics Project contract RII3-CT-2004-506078 and by the
Research Cooperation Program of Japan Society for the Promotion of Science
(JSPS) and Spanish CSIC.
\end{acknowledgments}

\end{document}